\documentstyle[12pt,epsfig]{article}
\date{}
\begin{document}
\def\bt{\begin{tabular}}
\def\et{\end{tabular}}
\def\bfr{\begin{flushright}}
\def\mm{\mbox{\boldmath $ }}
\def\efr{\end{flushright}}
\def\bfl{\begin{flushleft}}
\def\efl{\end{flushleft}}
\def\vs{\vspace}
\def\hs{\hspace}
\def\sta{\stackrel}
\def\pb{\parbox}
\def\bc{\begin{center}}
\def\ec{\end{center}}
\def\sp{\setlength{\parindent}{2\ccwd}}
\def\bp{\begin{picture}}
\def\ep{\end{picture}}
\def\uni{\unitlength=1mm}
\def\REF#1{\par\hangindent\parindent\indent\llap{#1\enspace}\ignorespaces}

{\sf
\title{
{\sf Iterative Solution for Generalized Sombrero-shaped Potential in
$N$-dimensional Space\thanks{Work supported in part by NNSFC under
No. 10547001}}}
\author{
W. Q. Zhao\\
{\small \it  China Center of Advanced Science and Technology (CCAST)}\\
{\small \it         (World Lab.), P.O. Box 8730, Beijing 100080,  China}\\
{\small \it  Institute of High Energy Physics, Chinese Academy of
Sciences, Beijing 100039, China}}
 \maketitle
%\date{}

 \vspace{4cm}

\begin{abstract}
{\sf An explicit convergent iterative solution for the lowest energy
state of the Schroedinger equation with generalized $N$-dimensional
Sombrero-shaped potential is presented. The condition for the
convergence of the iteration procedure and the dependence of the
shape of the groundstate wave function on the parameters are
discussed.}
\end{abstract}

\vspace{.5cm}

PACS{:~~11.10.Ef,~~03.65.Ge}\\


}
\newpage

{\large \sf

\section*{\bf 1. Introduction}
\setcounter{section}{1} \setcounter{equation}{0}

In this paper, the iterative method of Friedberg, Lee and Zhao[1]
will be applied to a generalized radially symmetric Sombrero-shaped
potential in $N$-dimensional space with
$$
V(r) = \frac{1}{2} g^2 (r^2-r_0^2)^2(r^2+Ar_0^2),\eqno(1)
$$
where $r_0^4=(2+N)/3$, $g^2$ and $A$ are arbitrary constants. The
same potential in the special case of $N=1$ has been studied
recently[2]. As noted in Ref.[2] this investigation is stimulated by
an interesting question raised by Roman Jackiw[3]. For arbitrary
$N$, the corresponding Schroedinger equation for the groundstate
radial wave function is
$$
(-\frac{1}{2r^{2k}}\frac{d}{dr}r^{2k}\frac{d}{dr}+V(r))\psi(r)=E\psi(r)\eqno(2)
$$
with $ k=(N-1)/2$. The boundary conditions are
$$
\psi(\infty)=0~{\sf and}~\psi'(0)=0.\eqno(3)
$$
When $g=1$ and $A=2$ the solution of the groundstate has an
analytical form as $\psi(r)=e^{-r^4/4}$ with the eigenvalue
$E_0=r_0^6$. However, for arbitrary $g$ and $A$ the groundstate wave
function has no analytical form.

To apply the iterative method we introduce a trial function
$\phi(r)$ satisfying another Schroedinger equation
$$
(-\frac{1}{2r^{2k}}\frac{d}{dr}r^{2k}\frac{d}{dr}+V(r)-h(r))\phi(r)
=gE_0\phi(r)=(E-\Delta)\phi(r),\eqno(4)
$$
where $h(r)$ and $\Delta$ are the corrections of the potential and
the groundstate energy. The explicit construction of the trial
function $\phi(r)$ will be given in the next section. Here, we note
some useful relations.

Multiplying (4) on the left by $\psi(r)$ and (2) by $\phi(r)$, their
difference gives
$$
-\frac{1}{2r^{2k}}\frac{d}{dr}(\psi r^{2k}\frac{d}{dr}\phi-\phi
r^{2k}\frac{d}{dr}\psi)=(h-\Delta)\phi\psi.\eqno(5)
$$
Let
$$
\psi(r) = \phi(r)f(r).\eqno(6)
$$
The equations for $f(r)$ and $\Delta$ can be deduced by using
$$
\frac{d}{dr}\bigg[r^{2k}\phi^2\frac{df}{dr}\bigg]=2r^{2k}(h-\Delta)\phi^2f.\eqno(7)
$$
The integration of the left-hand side of (7) over $r=0$ to $\infty$
is zero. This gives the expression of the energy correction
$$
\Delta=\frac{\int\limits_0^{\infty}r^{2k}\phi^2(r)
h(r)f(r)dr}{\int\limits_0^{\infty} r^{2k}\phi^2(r)f(r)dr}.\eqno(8)
$$
Introducing two iterative series $\{f_n(r)\}$ and $\{\Delta_n\}$
with $n=0,~1,~\cdots$, we obtain the following iteration equations
$$
\Delta_n=\frac{\int\limits_0^{\infty}r^{2k}\phi^2(r)
h(r)f_{n-1}(r)dr}{\int\limits_0^{\infty}
r^{2k}\phi^2(r)f_{n-1}(r)dr},\eqno(9)
$$
$$
f_n(r)=f_n(r_c)-2\int\limits_{r_c}^{r} \frac{dy}{y^{2k}\phi^2(y)}
\int\limits_{r_c}^{y}x^{2k}\phi^2(x)
(\Delta_n-h(x))f_{n-1}(x)dx,\eqno(10)
$$
where $r_c$ could be chosen as $r_c=0$ or $r_c=\infty$. To ensure
the convergency of the iterative method it is necessary to construct
the trial function in such way that the perturbed potential $h(r)$
is always positive (or negative) and finite everywhere. Specially,
$h(r)\rightarrow 0$ when $r\rightarrow \infty$. In the following we
construct the trial function for the iteration procedure. As we
shall see, in the cases that we have examined the iterations give
rapidly convergent results.

\newpage

\section*{\bf 2. Trial Functions}
\setcounter{section}{2} \setcounter{equation}{0}

We first introduce a trial function
$$
\phi_+(r)=\bigg(\frac{r_0+a}{r+a}\bigg)^k e^{-g{\cal S}_0(r)-{\cal
S}_1(r)}\eqno(11)
$$
satisfying the following Schroedinger equation
$$
(-\frac{1}{r^{2k}}\frac{d}{dr}r^{2k}\frac{d}{dr}+V(r)-h_+(r))\phi_+(r)
=gE_0\phi_+(r)\eqno(12)
$$
and the boundary condition
$$
\phi_+(\infty)=0.\eqno(13)
$$
Substituting (11) into (12), we compare terms with the same power of
$g$. From $g^2$-terms we obtain
$$
{\cal S}_0'(r)=\sqrt{2v}=(r^2-r_0^2)\sqrt{r^2+A r_0^2}.\eqno(14)
$$
To ensure $h_+(r)$ satisfying the convergence condition, $S_1(r)$
is defined in a special way to prevent terms with positive powers
of $r$ presenting in $h_+(r)$. For $g^1$ terms we have
$$
-\frac{1}{2}\bigg(2{\cal S}_0'{\cal S}_1'-{\cal
S}_0''-2ka\sqrt{r^2+Ar_0^2}\bigg)\bigg|_{r=r_0}-ka^2=E_0.\eqno(15)
$$
Introducing
$$
E_0=E_0^{(1)}+E_0^{(2)}+E_0^{(3)}\eqno(16)
$$
and defining
$$
E_0^{(3)}=-ka^2\eqno(17)
$$
we write
$$
{\cal S}_0'{\cal S}_1'=(\frac{1}{2}{\cal
S}_0''-E_0^{(1)})+(ka\sqrt{r^2+Ar_0^2}-E_0^{(2)}).
$$
Since $S_0'(r_0)=0$ we obtain
$$
E_0^{(1)}=\frac{1}{2}{\cal S}_0''(r_0)=r_0^2\sqrt{1+A},\eqno(18)
$$
$$
E_0^{(2)}=kar_0\sqrt{1+A}\eqno(19)
$$
and
$$
{\cal S}_1'=(\frac{1}{2}{\cal S}_0''-E_0^{(1)})/{\cal S}_0'
+(ka\sqrt{r^2+Ar_0^2}-E_0^{(2)})/{\cal S}_0'.\eqno(20)
$$
Substituting ${\cal S}_0'(r)$ into (20) we have explicitly
$$
{\cal S}_1'(r)=\frac{r(3r^2+2Ar_0^2-r_0^2)
-2r_0^2\sqrt{1+A}\sqrt{r^2+Ar_0^2}}{2(r^2-r_0^2)(r^2+Ar_0^2)}
$$
$$
+\frac{ka}{\sqrt{r^2+Ar_0^2}(\sqrt{r^2+Ar_0^2}+r_0\sqrt{1+A})}.\eqno(21)
$$
The expression for $h_+(r)$ is
$$
h_+(r)=\frac{1}{2}({\cal S}_1'^2-{\cal
S}_1'')+\frac{1}{2}~\frac{k(k+1)}{(r+a)^2} -\frac{ka}{r(r+a)}{\cal
S}_1'-\frac{k^2}{r(r+a)}
$$
$$
+kag(r_0^2-a^2)\frac{\sqrt{r^2+Ar_0^2}}{r(r+a)}
+ka^2g\frac{Ar_0^2}{r(\sqrt{r^2+Ar_0^2}+r)}.\eqno(22)
$$
The condition $ \phi'(0)=0$ can be satisfied by introducing the
trial function as
$$
\phi(r)=\phi_+(r)+\xi \phi_-(r)~~{\sf for}~~r<r_0\eqno(23a)
$$
and
$$
\phi(r)=\bigg(1+\xi \phi_-(r_0)/\phi_+(r_0)\bigg)\phi_+(r)~~{\sf
for}~~r>r_0\eqno(23b)
$$
where $\phi_-(r)$ is defined as
$$
\phi_-(r)=\bigg(\frac{r_0+a}{r+a}\bigg)^k e^{-g{\cal S}_0(-r)-{\cal
S}_1(r)}.\eqno(24)
$$
The parameter $\xi$ is fixed to satisfy the condition
$\phi'(0)=0$, namely
$$
\phi'_+(0)+\xi \phi'_-(0)=0.\eqno(25)
$$
Correspondingly $\phi(r)$ satisfies the Schroedinger equation (4)
with
$$
h(r)=h_+(r)~~{\sf for}~~r>r_0~~\eqno(26)
$$
and
$$
h(r)=h_+(r)+2g\xi\bigg(E_0+ka\frac{a^2-r_0^2}{r(r+a)}\sqrt{r^2+Ar_0^2}
$$
$$
-ka^2\frac{Ar_0^2}{r(\sqrt{r^2+Ar_0^2}+r)}\bigg)\phi_-(r)/\phi(r)\eqno(27)
$$
for $r<r_0$. It is interesting to notice that the condition (25) for
$\phi'(0)=0$ also ensures $h(r)$ to be finite when $r\rightarrow 0$,
which is necessary for the convergency of the iteration procedure.

By integrating (14) and (21) we obtain ${\cal S}_0(r)$ and ${\cal
S}_1(r)$ as

$$
{\cal S}_0(r)=\frac{1}{8}r\sqrt{r^2+Ar_0^2}(2r^2+Ar_0^2-4r_0^2)
-\frac{1}{8}(A^2r_0^4+4Ar_0^4)\ln (r+\sqrt{r^2+Ar_0^2})\eqno(28)
$$
$$
{\cal S}_1(r)=\ln (r+r_0)+\frac{1}{4}\ln
(r^2+Ar_0^2)+(\frac{1}{2}+\frac{ka}{2r_0}) \ln
\frac{\sqrt{1+A}\sqrt{r^2+Ar_0^2}+r+Ar_0}
{\sqrt{1+A}\sqrt{r^2+Ar_0^2}-r+Ar_0}.\eqno(29)
$$
Substituting them into (12), (24) and (23a-b) gives the final
expression of the trial function. From (21) we can also reach
$$
\frac{1}{2}({\cal S}_1'^2-{\cal
S}_1'')=\frac{\gamma}{8(r^2+Ar_0^2)^2(\alpha+\beta)}
+\frac{ka}{2r(r+a)(r^2+Ar_0^2)}\frac{\gamma'}{\alpha'+\beta'}
$$
$$
+\frac{k^2a^2}{2(r^2+Ar_0^2)(\sqrt{r^2+Ar_0^2}+r_0\sqrt{1+A})^2}
$$
$$
+\frac{kar(2\sqrt{r^2+Ar_0^2}+r_0\sqrt{1+A})}{(r^2+Ar_0^2)^{3/2}(\sqrt{r^2+Ar_0^2}+r_0\sqrt{1+A})^2}
\eqno(30)
$$
where
$$
\gamma=\frac{\alpha^2-\beta^2}{(r^2-r_0^2)^2},~~~
\gamma'=\frac{\alpha'^2-\beta'^2}{(r^2-r_0^2)^2}\eqno(31)
$$
with
$$
\alpha=15r^6+(18A-6)r^4r_0^2+(8A^2+12A+7)r^2r_0^4+(8A^2+2A)r_0^6,
$$
$$
\beta=8\sqrt{1+A}r_0^2r\bigg(3r^2+(2A-1)r_0^2\bigg)\sqrt{r^2+Ar_0^2},
$$
$$
\alpha'=r(3r^2+(2A-1)r_0^2)
$$
and
$$
\beta'=2r_0^2\sqrt{1+A}\sqrt{r^2+Ar_0^2}.
$$
Explicitly
$$
\gamma=225r^8+270(1+2A)r^6r_0^2+3(188A^2+216A-5)r^4r_0^4
$$
$$
+36A(8A^2+10A-1)r^2r_0^6+4(16A^4+8A^2+1)r_0^8
$$
and
$$
\gamma'=9r^4+3(4A-1)r^2r_0^2+4A(1+A)r_0^4.
$$
Substituting (30) into (22) and (27) gives the final expression of
$h$. With above results for $h$ and $\phi$ we are ready to perform
the iteration.\\

\section*{\bf 3. Numerical Result}
\setcounter{section}{3} \setcounter{equation}{0}

Starting from the above defined trial function $\phi(r)$ and the
related $h(r)$, we can perform the iteration based on (9) and (10).
Our numerical results show that the finally obtained wave functions
and eigenvalues for the groundstate convergent nicely. Let us take
$N=3$ as an example.

For $g=1$ and $A=2$, the exact solution of the groundstate is
$$
\phi(x)=e^{-r^4/4}
$$
with $E_0=r_0^6$. Starting from the trial function defined by (23a)
and (23b), with arbitrarily chosen parameter $a$ and the
corresponding parameter $\xi$ fixed by (25), the iteration procedure
(9) and (10) gives the final convergent result of the wave function
and the eigenvalue of the groundstate, which is consistent to the
exact solution. The trial function and the convergent groundstate
wave function after the iteration are plotted in Fig. 1. It is
interesting to observe the transition of the shape of the wave
function for the trial function with maxima at a finite $r$ to the
final convergent one with only one maximum at $r=0$, as the exact
groundstate wave function should be. This answered the question
raised by R. Jackiw[3] in $N$-dimensional case: Even the trial
function proposed has its maxima at $r>0$ the iteration procedure
would still reach the exact solution of the groundstate wave
function with its only maximum at $r=0$.

As examples the obtained groundstate wave functions after the
iteration procedure are plotted in Figs.~2 and 3 for $A=2$ and
$g=0.5,~1$ or $2$, and for $g=1$ and $A=1,~2$ or $3$, respectively.
In Table 1 the corresponding eigenvalues of the groundstate obtained
from the iteration are listed for different parameters $g$ and $A$.
It can be seen that the iterative energy series convergent quite
nicely. From the figures it is interesting to see the transition of
the form of the obtained groundstate wave functions from the shape
with maximum at $r=0$ to the one with maxima at a finite $r$,
becoming a degenerate groundstate, when $g$ increases from $<1$,
passing $1$ to $>1$ for $A=2$, or when $A$ increases from $<2$,
passing $2$ to $>2$ for $g=1$. The results seem to show that the
groundstate wave functions in the region $g\leq 1$ and $A\leq 2$
have the shape with only one maximum at $r=0$, while in the region
outside the wave functions become degenerate at a finite $r$. Their
maxima move to larger $r$ when the parameters $g$ and $A$ increase
further.

\section*{\bf Acknowledgement}
\setcounter{section}{3} \setcounter{equation}{0}

The author would like to thank Professor T. D. Lee for his
continuous guidance and instruction, and for carefully revising the manuscripts.\\

\noindent {\bf References}

1. R. Friedberg, T. D. Lee and W. Q. Zhao, Ann. Phys. 321(2006)1981

~~~~R. Friedberg and T. D. Lee, Ann. Phys. 316(2005)44

2. R. Friedberg, T. D. Lee and W. Q. Zhao, arXiv: 0709.1997,

~~~~Ann. Phys. (2007)doi:10.1016/j.aop.2007.09.006

3. R. Jackiw, Private communication

\vspace{1cm}

Table 1. Eigenvalues of groundstates in the iteration for $N=3$ \\

\begin{tabular}{|c|c|c|c|c|c|c|}
  \hline
 $g$~~~~$A$ & $gE_0$ & $gE_1$ & $gE_2$ & $gE_3$ & $gE_4$ &$gE_5$\\
  \hline
 0.5~~2~& -0.4300 & 1.3963 & 1.3795 & 1.3775 & 1.3773 & 1.3772 \\
\hline
 1~~~~2 & -8.6479 & 2.1523 & 2.1516 & 2.1517 & 2.1517 & 2.1517 \\
  \hline
 2~~~~2 & 5.5581 & 4.1362 & 4.0976 & 4.1108 & 4.1092 & 4.1094\\
  \hline
 1~~~1 & -2.3537 & 1.8920 & 1.8473 & 1.8402 & 1.8393 & 1.8392 \\
\hline
  1~~~3 & 3.6773 & 2.4675 & 2.4353 & 2.4425 & 2.4417 & 2.4418 \\
  \hline
\end{tabular}

\begin{figure}[h]
 \centerline{
\epsfig{file=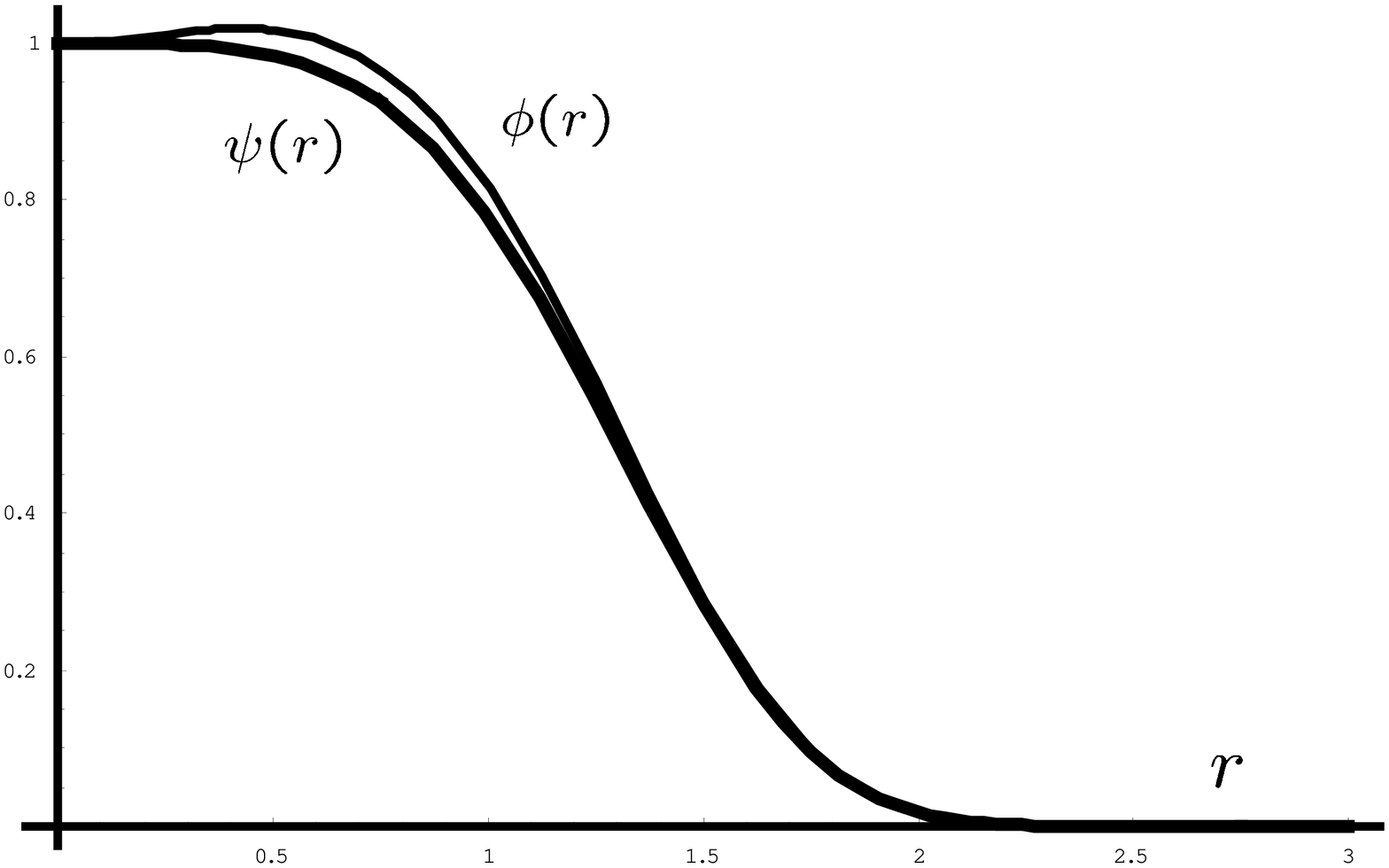, width=8cm, height=14cm, angle=-90}}
\vspace{.5cm}
 \centerline{{\normalsize \sf Fig. 1~  Trial Function $\phi(r)$ and Groundstate Wave
 Function $\psi(r)$ }}
\centerline{{\normalsize \sf for $N=3$, $g=1$ and $A=2$.}}
\end{figure}

\begin{figure}[h]
 \centerline{
\epsfig{file=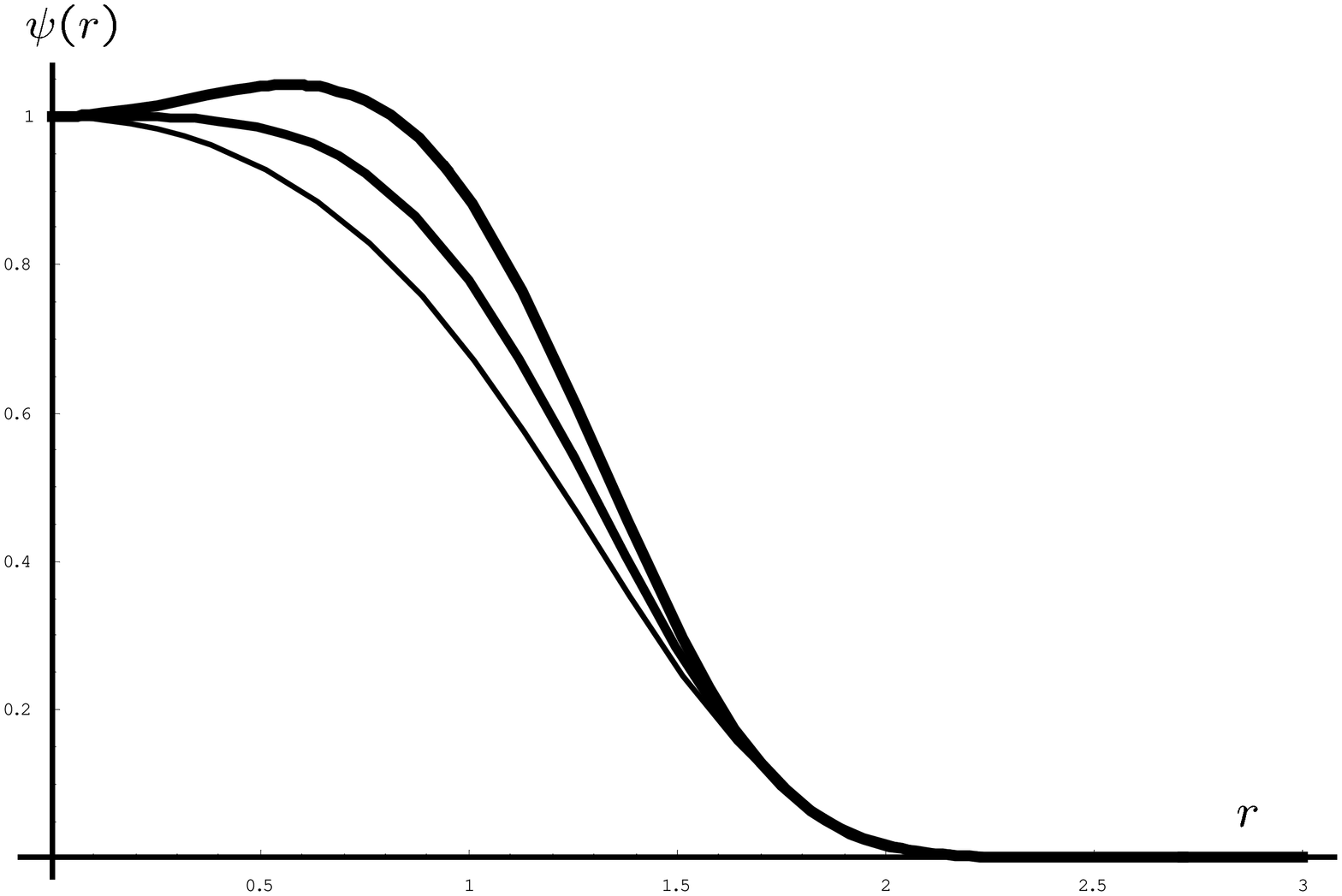, width=8cm, height=14cm, angle=-90}}
\vspace{.5cm}
 \centerline{{\normalsize \sf Fig. 2~  Groundstate Wave
 Function $\psi(r)$ for $N=3$, $g=1$ }}
 \centerline{{\normalsize \sf and $A=1$ (thin), $2$ (middle) and $3$ (thick).}}
\end{figure}

\begin{figure}[h]
 \centerline{
\epsfig{file=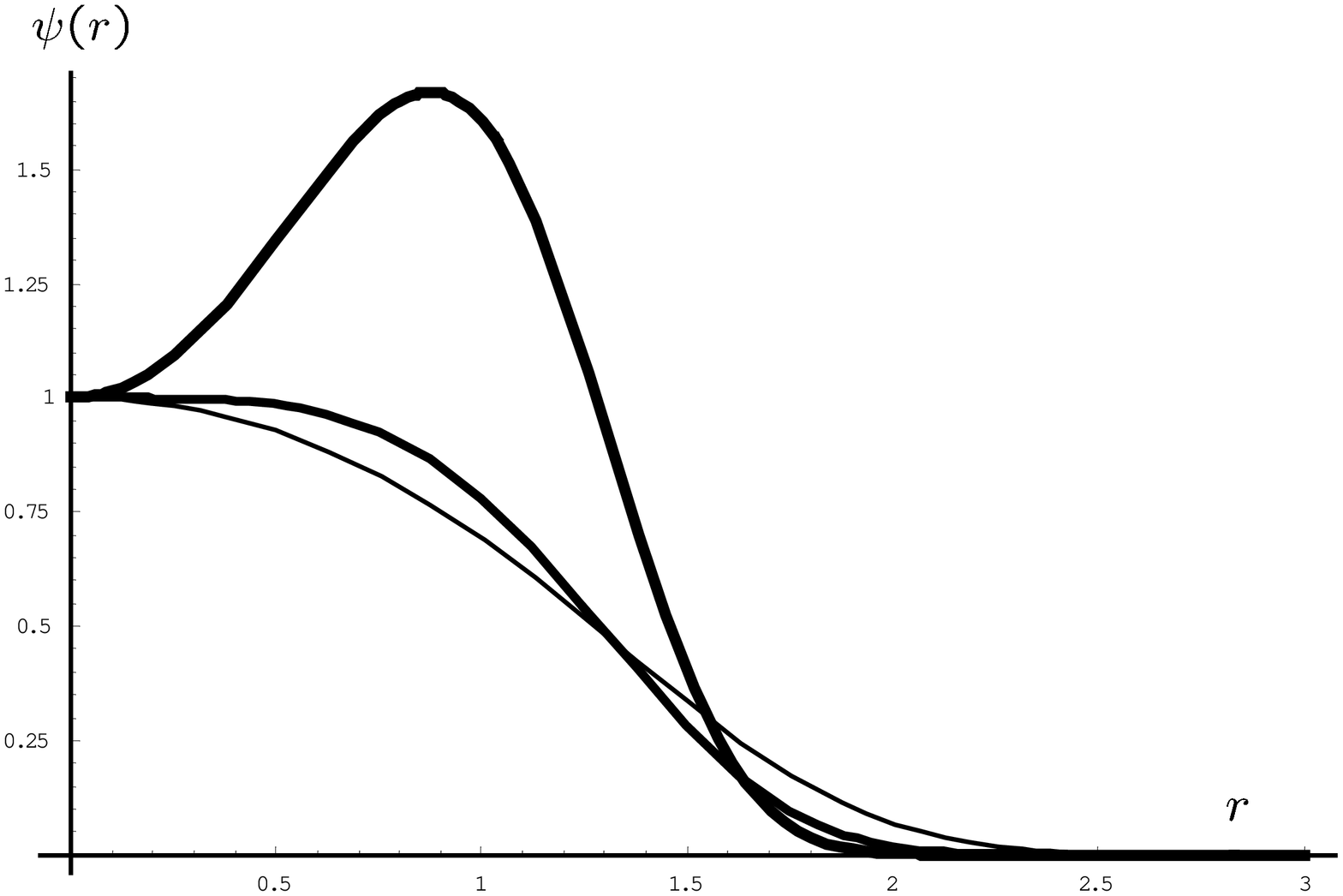, width=8cm, height=14cm, angle=-90}}
\vspace{.5cm}
 \centerline{{\normalsize \sf Fig. 3~  Groundstate Wave
 Function $\psi(r)$ for $N=3$, $A=2$ }}
 \centerline{{\normalsize \sf and $g=0.5$ (thin), $1$ (middle) and $2$ (thick).}}
\end{figure}

 }
\end{document}